\documentclass[conference]{IEEEtran}
\IEEEoverridecommandlockouts
\usepackage{cite}
\usepackage{amsmath,amssymb,amsfonts}
\usepackage{algorithmic}
\usepackage{graphicx}
\usepackage{textcomp}
\usepackage{xcolor}
\usepackage{algorithm}
\usepackage{algorithmic} 
\usepackage{subfigure}
\usepackage{parskip}
\usepackage{booktabs}
\usepackage{bbding}
\usepackage{caption}
\usepackage{indentfirst} 
\def\BibTeX{{\rm B\kern-.05em{\sc i\kern-.025em b}\kern-.08em
    T\kern-.1667em\lower.7ex\hbox{E}\kern-.125emX}}
\begin{document}

\title{Deep Joint Source-Channel Coding Based on Semantics of Pixels}
    \author{  Qizheng Sun,
  	Caili Guo,
  	Yang Yang,
  	Jiujiu Chen,
  	Rui Tang,
  	Chuanhong Liu\\
  	\IEEEauthorblockA{
  		Beijing University of Posts and Telecommunications,\\
  		Beijing Key Laboratory of Network System Architecture and Convergence,\\
  		Beijing Laboratory of Advanced Information Networks,
  		 100876, Beijing, China\\
  		Email: \{qizheng\_sun, guocaili, yangyang01, chenjiujiu, 1361680482, 2016\_liuchuanhong\}@bupt.edu.cn}
  	} 
       
\maketitle

\begin{abstract}
The semantic information of the image for intelligent tasks is hidden behind the pixels, and slight changes in the pixels will affect the performance of intelligent tasks.
In order to preserve semantic information behind pixels for intelligent tasks during wireless image transmission, we propose a joint source-channel coding method based on semantics of pixels, which can improve the performance of intelligent tasks for images at the receiver by retaining semantic information.
Specifically, we first utilize gradients of intelligent task's perception results with respect to pixels to represent the semantic importance of pixels. Then, we extract the semantic distortion, and train the deep joint source-channel coding network with the goal of minimizing semantic distortion rather than pixel's distortion. 
Experiment results demonstrate that the proposed method improves the performance of the intelligent classification task by 1.38\% and 66\% compared with the SOTA deep joint source-channel coding method and the traditional separately source-channel coding method at the same transmission rate and signal-to-noise ratio.
\end{abstract}


\section{Introduction}
\begin{figure}[t]
	\centering
	\includegraphics[width=0.8\linewidth]{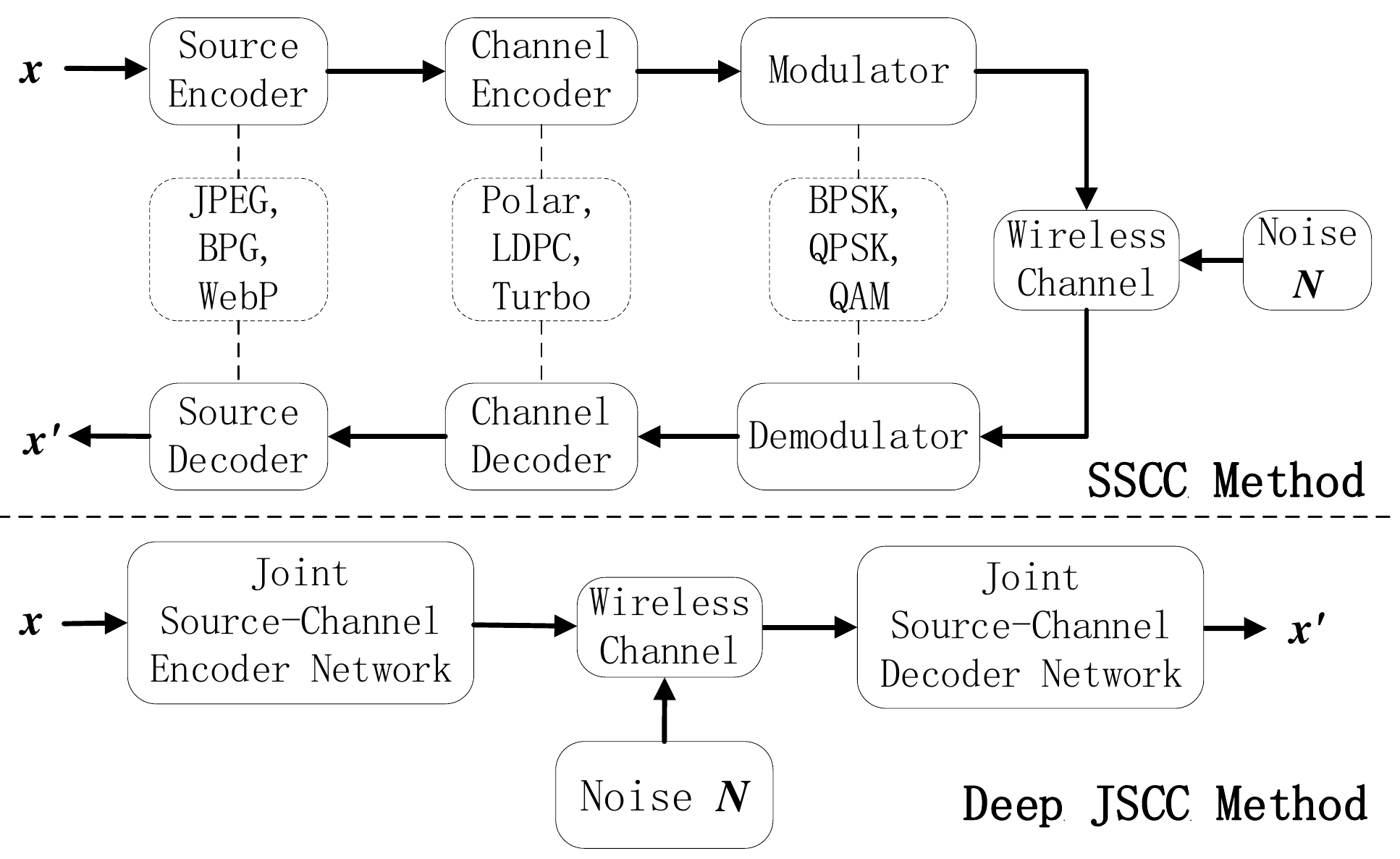}
	\caption{Block diagrams of wireless image transmission schemes. Top: SSCC method. Bottom: deep JSCC method.}    
	\vspace{-3ex}
	\label{intro_1}
\end{figure}

In the past few decades, the wireless mobile communication system represented by 5G has achieved great success. 
For image data, they employ the separate source-channel coding method (SSCC) under the guidance of Shannon's separation theorem. 
As shown in Fig \ref{intro_1}, first, SSCC uses the source coding algorithm (such as JPEG, WebP, BPG) to compress the redundant information, and then the source-independent channel coding method (such as LDPC, Polar, Turbo, etc.) is used to reconstruct in the presence of channel noise. 
When the source is infinite, SSCC can achieve optimal coding performance. 
However, in practice, the infinite bits' assumption of Shannon's separation theorem cannot be satisfied. In fact, the joint effects of source coding distortion and channel coding error affect the signal quality at the receiver \cite{yang2021deep}, so it is necessary to jointly consider source coding and channel coding. 
In addition, the rapid development of deep learning technology enables the joint consideration of source coding and channel coding. As shown in the Fig 1, deep learning-based joint source-channel coding (deep JSCC) refers to implement joint source encoding and channel encoding by end-to-end (E2E) semantic communication framework using deep neural networks.  Due to the powerful learning ability of deep neural networks, deep JSCC can learn how to remove the redundancy of source information and how to resist the channel noise \cite{yang2021deep}.

At the same time, 6G (sixth generation) puts forward the vision of smart interconnection of everything. On the one hand, the deep integration of communication and artificial intelligence (AI) is required, and research on semantic communication for intelligent tasks \cite{qin2021semantic} has become a trend. On the other hand, efficient coding methods are required.
Therefore, deep JSCC considering intelligent tasks has gained extensive attention. 
For image data, deep JSCC methods can be divided into two categories, one is task-oriented without reconstructing the image. In these works, only the semantic information related to the intelligent task is encoded, and the decoder directly performs the intelligent task \cite{jankowski2020wireless,liuchuanhong,wang2021deep,yang2021semantic}. 
However, this type of work only can directly perform the intelligent task, and cannot apply to the applications that require image reconstruction.
The other is that the receiver completes the image reconstruction. The encoder extracts the global semantic information, and the decoder reconstructs the image according to the received semantic information \cite{bourtsoulatze2019deep,kurka2020deep,yang2022deep,xu2021wireless}. 
However, existing deep JSCC methods for image reconstruction aim to optimize the visual quality. They only focus on the accurate transmission of pixel-level information, ignoring the semantic information required for downstream AI tasks.

The semantic information refers to that useful for serving the downstream AI task at the receiver \cite{qin2021semantic}, which directly affects the perception result and performance of the downstream AI task. 
There are some works have considered semantic information preserving during wireless image transmission. Wang et al.\cite{wang2022perceptual} used learned perceptual image patch similarity (LPIPS) \cite{zhang2018unreasonable} as loss function to yield images that are visually pleasing to humans. However, they focus on human visual perception without considering downstream AI tasks.
Shao et al. \cite{shao2021learning} leveraged an information bottleneck (IB) framework to formalize a rate-distortion trade off between the informativeness of the encoded feature and the inference performance in a task-oriented manner, i.e., targeting the downstream inference task rather than data reconstruction. Using IB can help to retain semantic information. However, the loss function of IB needs to use label information, so it is suitable for supervision tasks, but not suitable for reconstruction tasks. 
In addition, some studies in the field of image compression have considered content information \cite{li2018learning} and feature information \cite{yang2020discernible}.
However, on the one hand, semantic information is not utilized directly, and semantic consistency during transmission cannot be guaranteed. On the other hand, these studies only consider the source coding while ignore the channel coding, which cannot combat channel noise in the actual communication process.
In summary, these methods do not jointly consider communication tasks and the semantic information of downstream AI tasks in a stable and universal way.
Directly extracting the semantic information behind the pixels can avoid introducing errors due to complex design, and has better scalability. Thus, a semantics of pixels extraction method needs to be proposed for measuring semantic information.

Previous studies have shown that the relationship between the semantic information and the pixel's information is not strictly linear \cite{luo2022frequency}. 
In the field of neural network's interpretability, it can be shown that pixels of the object part are more important than the background part by visualizing the heatmap results \cite{selvaraju2017grad}. In the field of adversarial attacks, even an attack on a very small number of important pixels can produce completely different perceptual results \cite{cisse2017houdini}. All of these indicate that the importance of pixels to the perceptual results is different, that is, the semantic importance is different.
That means keeping the accurate transmission of pixel-level information does not guarantee the correct understanding of downstream AI tasks. 

Motivated by this, the deep joint source-channel coding based on semantics of pixels (SP-JSCC) method is proposed for wireless image transmission. In this paper, SP-JSCC designs a semantic distortion extractor to preserve semantic information at the receiver by minimizing the semantic distortion. we quantify the semantic importance of pixels through the idea of gradients, which can represent the contribution of pixels to perception results. Then, we design the semantic distortion extractor, which is the core of SP-JSCC and can extract the semantic distortion at the transmitter and the receiver. We use the semantic distortion as the loss function to train the deep joint source-channel codec network in an end-to-end manner, which can retain the semantic information needed by the downstream AI task and so as to improve the task performance of images at the receiver.

\section{System Model and Problem Description}

\subsection{System Model}
\begin{figure}[t]
	\centering
	\includegraphics[width=0.9\linewidth]{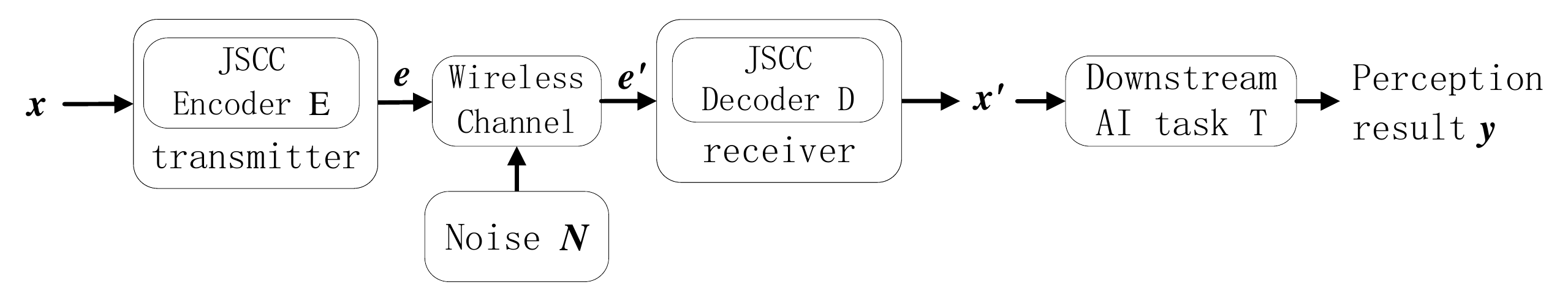}
	\caption{JSCC system model diagram considering AI tasks.}	
	\vspace{-3ex}
	\label{intro_2}
\end{figure}
Fig \ref{intro_2} shows an end-to-end communication system for wireless image transmission considering the downstream AI task. An input image $\textbf{\textit{x}} \in \mathcal{R}^{\textit{n}}$ of dimension $\textit{n}$ is to be transmitted, where $\mathcal{R}$ denotes the set of real numbers, and the transmitter maps the input image $\textbf{\textit{x}}$ into a complex-valued symbolic vector $\textbf{\textit{e}}$ after the JSCC encoder $\mathrm{E}$, which can be expressed as:
\begin{equation}
\textbf{\textit{e}}=\mathrm{E}\left(\textbf{\textit{x}}, \theta_{1}\right) \in \mathcal{C}^{\textit{s}},
\end{equation}
where $\textit{s}$ denotes the dimension of $\textbf{\textit{e}}$, $\mathcal{C}$ denotes the set of complex numbers, and $\theta_{1}$ denotes the parameters of the JSCC encoder $\mathrm{E}$. The symbol vector $\textbf{\textit{e}}$ after encoding is transmitted over a noisy AWGN channel, which can be expressed as:
\begin{equation}
\textbf{\textit{e}}^{\prime}=\textbf{\textit{e}}+\textbf{\textit{N}} \in \mathcal{C}^{\textit{s}},
\end{equation}
where $\textbf{\textit{N}} \in \mathcal{C}^{\textit{s}}$ denotes the noise of the channel. 
The receiver map $\textbf{\textit{e}}^{\prime}$ to the reconstructed image $\textbf{\textit{x}}^{\prime}$ through the JSCC decoder $\mathrm{D}$, which can be expressed as:
\begin{equation}
\textbf{\textit{x}}^{\prime}=\mathrm{D}\left(\textbf{\textit{e}}^{\prime}, \theta_{2}\right)=\mathrm{D}\left(\mathrm{E}\left(\textbf{\textit{x}}, \theta_{1}\right)+\textbf{\textit{N}}, \theta_{2}\right),
\end{equation}
where the reconstructed image $\textbf{\textit{x}}^{\prime} \in \mathcal{R}^{\textit{n}}$ is an estimate of the original image $\textbf{\textit{x}}$, and $\theta_{2}$ is the parameter of the JSCC decoder $\mathrm{D}$. Then, the reconstructed image $\textbf{\textit{x}}^{\prime}$ is passed through the downstream AI task $\mathrm{T}$, and perception results are obtained, which can be expressed as:
\begin{equation}\label{eq4}
\textbf{\textit{y}}=\mathrm{T}\left(\textbf{\textit{x}}^{\prime}\right),
\end{equation}
where $\textbf{\textit{y}}=\left[\textit{y}^{1}, \textit{y}^{2}, \ldots, \textit{y}^{C}\right]$, $\textit{y}^{c}(c \in\{1, \ldots, C\})$ denotes the $\textit{c}$-th perception result and $C$ is the total number of perception results.
\subsection{Problem Description}
The existing deep JSCC methods  \cite{bourtsoulatze2019deep,kurka2020deep,yang2022deep,xu2021wireless} use the pixel-level difference between $\textbf{\textit{x}}^{\prime}$ and  $\textbf{\textit{x}}$ as the loss function to train the joint source-channel codec network, which can be expressed as:
\begin{equation}
\mathcal{L}_{\text {deep JSCC }}=\textit{d}\left(\textbf{\textit{x}}, \textbf{\textit{x}}^{\prime}\right)=\left\|\textbf{\textit{x}}-\textbf{\textit{x}}^{\prime}\right\|^{2}.
\end{equation}
$\mathcal{L}_{\text {deep JSCC }}$ enables $\textbf{\textit{x}}^{\prime}$ to obtain a clear visual quality that is close to the original image $\textbf{\textit{x}}$. This approach only maintains the pixel-level consistency during image transmission without considering the perception results of the downstream AI task.

In addition, Wang et al.\cite{wang2022perceptual} use learned perceptual image patch similarity (LPIPS) \cite{zhang2018unreasonable} as the distortion part of loss function:

\begin{equation}
\mathcal{L}_{\text {LPIPS }}=\textit{d}_\text{LPIPS}\left(\textbf{\textit{x}}, \textbf{\textit{x}}^{\prime}\right).
\end{equation}
But it targets the clarity of human vision and does not consider the performance of downstream AI tasks.


Since the relationship between the semantic information and the pixel's information is not strictly linear, studying JSCC that preserves the semantic information of downstream AI tasks during image transmission is meaningful. 

\section{SP-JSCC Method}
\begin{figure}[t]
	\centering
	\includegraphics[width=0.9\linewidth]{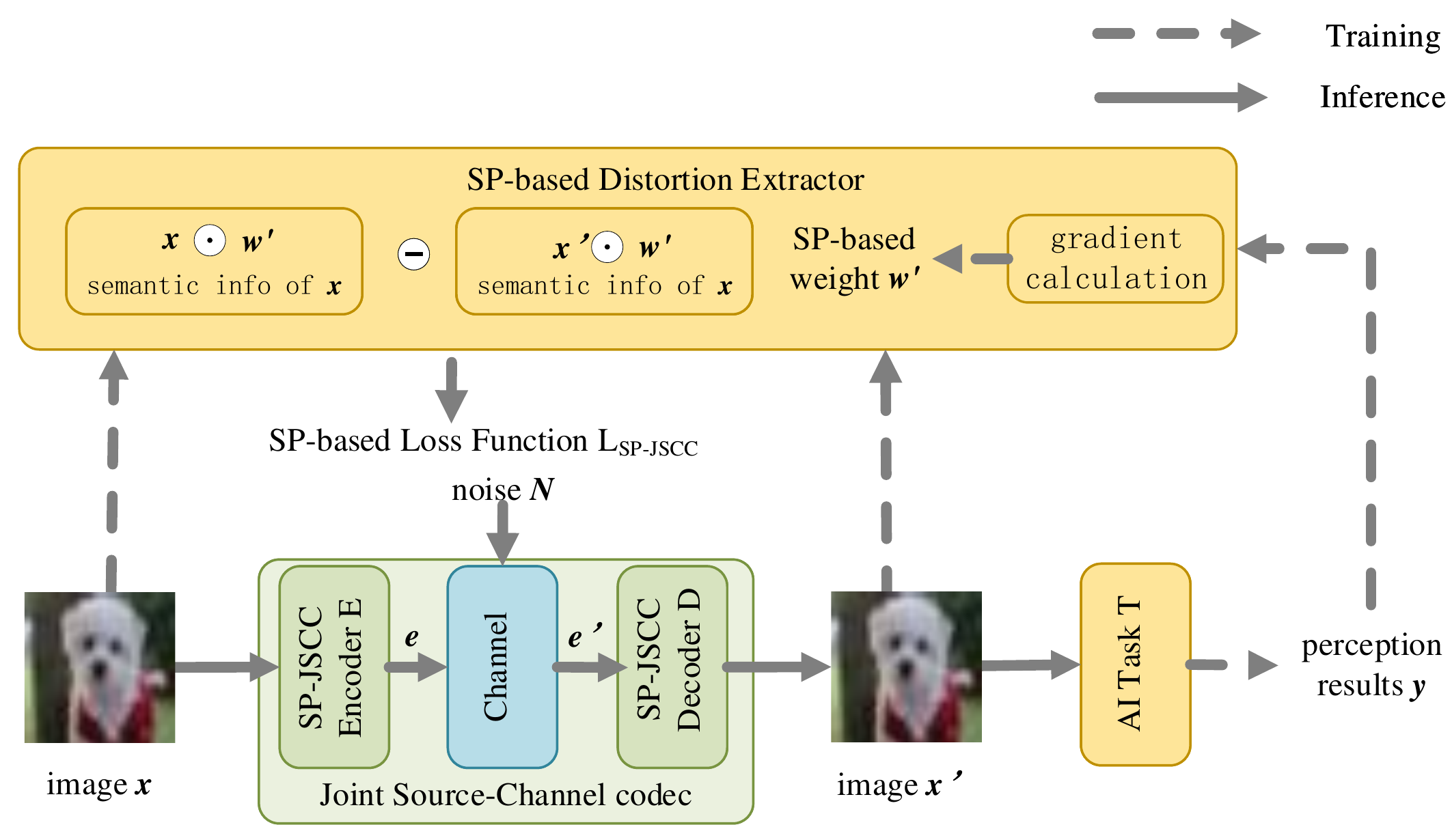}
	\caption{SP-JSCC method architecture diagram. }    
	\label{arch_1}
\end{figure}

\begin{figure}[t]
	\centering
	\includegraphics[width=0.9\linewidth]{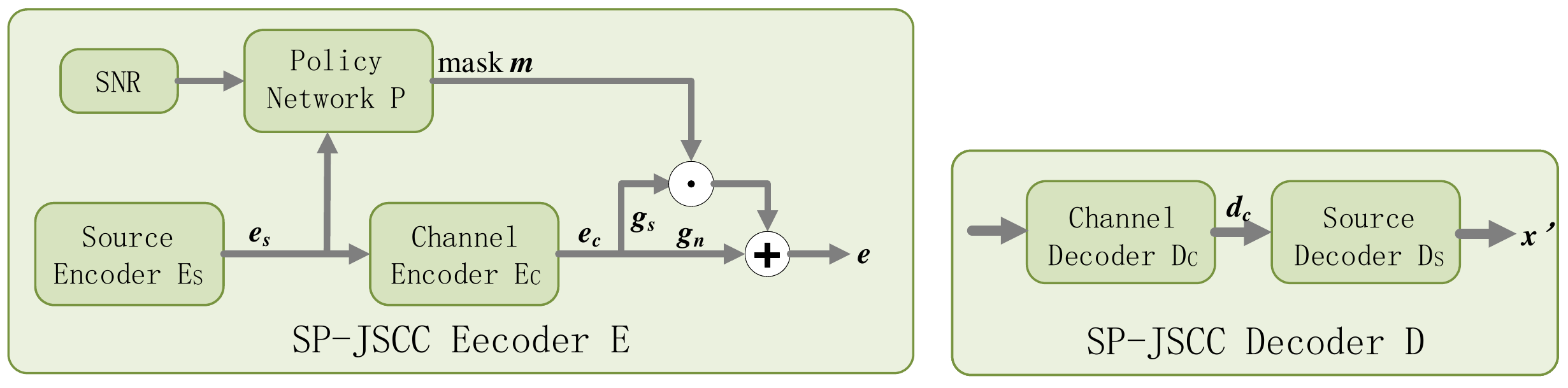}
	\caption{Detail design diagram of SP-JSCC codec.}    
	\vspace{-3ex}
	\label{arch_2}
\end{figure}
The semantic information of an image is extracted from its pixels, and thus each pixel contains a certain-level semantic information. That means, different pixel may have a different importance level to the downstream AI tasks \cite{goodfellow2014explaining}. Inspired by this, we design the joint source-channel coding based semantics of pixels method, which utilizes the importance degree of pixels to preserve the semantic information behind pixels of the downstream AI task. 

In the field of adversarial attacks, there are many methods utilize the gradient information of the perception result, such as Fast Gradient Sign Method (FGSM) \cite{goodfellow2014explaining} and Houdini \cite{cisse2017houdini}, which shows that the gradient information can effectively represent the importance degree. Inspired by this, we use gradients of perception results with respect to pixels as semantic importance degree of pixels for the downstream AI task. 

As shown in Fig \ref{arch_1}, SP-JSCC uses the perception results of the downstream AI task to calculate semantics of pixels based distortion (SP-based distortion) as loss function, and use it to guide the joint source-channel codec module. SP-based distortion extractor is the core of SP-JSCC method. Specifically, first, we calculate the semantics of pixels based weights (SP-based weights) $\textbf{\textit{W}}^{\prime}$ and use it to characterize the SP-based distortion between images at the transmitter $\textbf{\textit{x}}^{\prime}$ and receiver $\textbf{\textit{x}}$ as SP-based loss function. Then, we use the SP-based loss function to train the joint source-channel codec. 

\subsection{SP-based Weights Calculation}
This section gives the formula representation of the SP-based weights, which can facilitate the representation of subsequent SP-based distortion, namely SP-based loss function $\mathcal{L}_{\text {SP-JSCC }}$. SP-based weights can represent the importance degree of pixels to perception results, since it use gradients to calculate.

As shown in Fig \ref{arch_1}, we first calculate SP-based weights $\textbf{\textit{W}}^{\prime}$. Then use $\textbf{\textit{W}}^{\prime}$ to weight pixels to obtain the SP-based loss function.

First, we pre-train the downstream AI task's network with parameter ${\theta _0}$, and fixed ${\theta _0}$ in the following operation. Then, we complete inference with parameters ${\theta _0}$ to get perception results $\textbf{\textit{y}}$. The gradients of perception results with respect to pixels of image \textbf{\textit{x}} is used to quantify the semantic importance, which can be expressed as :
\begin{equation}
\textbf{\textit{w}}^{\textbf{\textit{c}}}= \frac{\partial \textit{y}^{\textit{c}}}{\partial \textbf{\textit{x}}},
\end{equation}
where $\textit{y}^\textit{c}$ is $\textit{c}$-th perception result. $\textbf{\textit{w}}^{\textbf{\textit{c}}}(\textit{k} \in\{1,2, \ldots, \textit{C}\})$ represents the semantic importance to the $\textit{c}$-th perception result $\textit{y}^\textit{c}$. The dimension of $\textbf{\textit{w}}^{\textbf{\textit{c}}}$ is the same as that of $\textbf{\textit{x}}$. To evaluate the comprehensive influence of $\textbf{\textit{x}}$ on all perception results $\textbf{\textit{y}}$, the average value of  $\textbf{\textit{w}}^{\textbf{\textit{c}}}$ over all perception results is calculated, which can be expressed as:
\begin{equation}
\textbf{\textit{w}}=\frac{1}{\textit{C}} \sum_{\textit{c}=1}^{\textit{C}} \textbf{\textit{w}}^{\textbf{\textit{c}}},
\end{equation}
whose dimension is also consistent with $\textbf{\textit{x}}$.

However, $\textbf{\textit{w}}$ cannot be directly used to design the loss function since it has negative values and its variance is too large. In particular, negative values of $\textbf{\textit{w}}$ will cause confusion in loss function, and large variance will cause the missing of some the pixel's information in loss function. Therefore, we post-process $\textbf{\textit{w}}$, and map $\textbf{\textit{w}}$ to $\textbf{\textit{w}}^{\prime}$ by:
\begin{equation}
\textbf{\textit{w}}^{\prime} = \text {L2\_norm}\left ( \left | \textbf{\textit{w}} \right |  \right ) ,
\end{equation}
where L2\_norm is the normalization using the L2 norm.

Note that since the design of the SP-based distortion extractor, the SP-based weights are specific to the downstream AI task, and therefore SP-JSCC is also task-specific. 

\subsection{SP-based Loss Function}
We introduce this section to formulate the SP-based loss function
$\mathcal{L}_{\text {SP-JSCC }}$ for wireless image transmission, which can extract the semantic distortion for the downstream AI task.

We use the architecture of JSCC with adaptive rate control \cite{yang2022deep}, which can support multiple rates using a single deep neural network (DNN) model and learn to dynamically control the rate based on the channel condition and image contents. 
As shown in Fig \ref{arch_2}, SP-JSCC encoder $\text {E}$ consists of the source encoder $\text {E}_{\text{S}}$, the channel encoder $\text {E}_{\text{C}}$ and the policy network ${\text {P}}$. SP-JSCC decoder consists of source decoder $\text {D}_{\text{S}}$ and channel encoder $\text {D}_{\text{C}}$. 
The SP-JSCC encoder $\text {E}$ is designed to achieve adaptive rate control. In particular, first, the source encoder $\text {E}_{\text{S}}$ extracts the image features $\textbf{\textit{e}}_{\textbf{\textit{S}}}$ from a source image $\textbf{\textit{x}}$. Then $\textbf{\textit{e}}_{\textbf{\textit{S}}}$ is fed to the channel encoder $\text {E}_{\text{C}}$ which generates $\left ( \textbf{\textit{g}}_{\textbf{\textit{s}}}  + \textbf{\textit{g}}_{\textbf{\textit{n}}} \right )$.  $\textbf{\textit{g}}_{\textbf{\textit{s}}}$ are the selective features that can be either active or inactive according to the policy network, whereas $\textbf{\textit{g}}_{\textbf{\textit{n}}}$ are non-selective features that are always active. The Policy network ${\text {P}}$ used to generate $\textbf{\textit{m}}$ according to the channel environment and signal content, and $\textbf{\textit{m}}$ is used to determine whether $\textbf{\textit{g}}_{\textbf{\textit{s}}}$ is activated. The encoded features $\textbf{\textit{e}}$ can be expressed as :
\begin{equation}
\textbf{\textit{e}}= \left ( \textbf{\textit{g}}_{\textbf{\textit{s}}} \cdot \textbf{\textit{m}} \right )  + \textbf{\textit{g}}_{\textbf{\textit{n}}}.
\end{equation}

The SP-based loss function is calculated using the semantic weights $\textbf{\textit{w}}^{\prime}$ to weight pixels of the image, which can be expressed as:
\begin{equation}
\mathcal{L}_{\text {SP-JSCC }}\left(\theta_{1}, \theta_{2}\right)= \textbf{\textit{w}}^{\prime} \times\left\|\textbf{\textit{x}}^{\prime}-\textbf{\textit{x}}\right\|^{2}.
\end{equation}
As shown in Fig \ref{arch_1}, first, the image is encoded by SP-JSCC encoder $\text{E}$. Then the noise in the wireless channel is simulated. Finally, we use SP-JSCC decoder $\text{D}$ to decode and recover the image. 
The end-to-end wireless image transmission process is jointly trained using the SP-based loss function $\mathcal{L}_{\text {SP-JSCC }}$ so that the semantic information beneficial to the downstream AI task is retained.

\begin{algorithm}[t]
	\caption{SP-JSCC method}
	\label{algorithm1}
	\textbf{Input}: An image dataset $\left\{{{\boldsymbol{x}}^1},...,{{\boldsymbol{x}}^n}\right\}$ with $n$ images.\\
	\textbf{Parameter}: SP-JSCC encoder parameter ${\theta _1}$, SP-JSCC decoder parameter ${\theta _2}$.\\
	\textbf{Output}: SP-JSCC based wireless image transmission system.
	\begin{algorithmic}[1] 
		\STATE Pre-train the downstream AI task's network with parameter ${\theta _0}$, and fixed ${\theta _0}$ in the following operation.
		\STATE Obtain the SP-based weights $\textbf{\textit{w}}^{\prime}$.
		\STATE Initialize the encoder and decoder's parameters ${\theta _1}$, ${\theta _2}$.
		\WHILE{ not converged }
		\STATE Image Encoding: $\textbf{\textit{e}} \leftarrow E({\theta _1},\textbf{\textit{x}})$.
		\STATE Noise channel: $\textbf{\textit{e}}^{\prime} \leftarrow \textbf{\textit{N}} + \textbf{\textit{e}}$.
		\STATE Image decoding: 
		$\textbf{\textit{x}}^{\prime} \leftarrow D({\theta _2},\textbf{\textit{e}}^{\prime})$.
		\STATE Calculate the semantic loss function $\mathcal{L}_{\text {SP-JSCC}}$.
		\STATE Update ${\theta _1}$ and ${\theta _2}$ according to $\mathcal{L}_{\text {SP-JSCC}}$.
		\ENDWHILE
		\STATE \textbf{return} The optimal model.
	\end{algorithmic}
\end{algorithm}

We summarize the steps of SP-JSCC as Algorithm 1. In step
1, we pre-train the downstream AI task's network. In step 2, we obtain the SP-based weights. In step 3-8, we compute the semantic loss function $\mathcal{L}_{\text {SP-JSCC}}$. In step 9, we use  $\mathcal{L}_{\text {SP-JSCC}}$ to train the joint source-channel codec network.

\section{Experiment}

\subsection{Evaluation Metrics}
This section evaluates the distortion of semantic information, the distortion of the pixel, and the transmission rate, respectively.

To evaluate the semantic distortion, we use the downstream classification task's performance at the receiver, including accuracy (ACC) and F1-score, where F1-score is the harmonic mean of precision and recall.
To evaluate the distortion of pixels, we use the peak signal-to-noise ratio (PSNR) and the structural similarity index measure (SSIM).
To evaluate the transmission rate, we use
wireless channel usage per pixel (CPP). Suppose the width and height of an input image are $\textit{W}$, $\textit{H}$ for RGB (3 channel) pixels respectively. The CPP is defined as $CPP=  \frac{\mathrm{L}\left [ \left ( \textbf{\textit{g}}_{\textbf{\textit{s}}} \cdot \textbf{\textit{m}} \right )  + \textbf{\textit{g}}_{\textbf{\textit{n}}} \right ] }{2\textit{H}\textit{W}}$, where the function $\text{L}$ represents the length.
The range of $CPP$ is $\left [ \frac{\mathrm{L} \left ( \textbf{\textit{g}}_{\textbf{\textit{n}}}\right ) }{2\textit{H}\textit{W}}, \frac{\mathrm{L} \left ( \textbf{\textit{g}}_{\textbf{\textit{s}}} + \textbf{\textit{g}}_{\textbf{\textit{n}}}\right ) }{2\textit{H}\textit{W}} \right ]$ depending on the mask $\textbf{\textit{m}}$ obtained by the policy network ${\text {P}}$. The 1/2 factor in CPP is because of complex-valued (quadrature) transmission over the wireless channel. 


\subsection{Implementation Details}
To serve as the baseline, we use JSCC with adaptive rate control (AR-JSCC) proposed by \cite{yang2022deep}. We also use BPG as the source coding method and LDPC as the channel encoding mode with regard to SSCC. LDPC codes is (1458, 1944), corresponding to rate 3/4. 16-QAM is used as the modulation method.
We evaluate these methods on the CIFAR-10 dataset which consists of 50000 training and 10000 testing images with 32$\times$32 pixels.


For channel, we consider the AWGN wireless channel. It is worth mentioning that we insert the additional SNR-adaptive modules \cite{xu2021wireless} between layers. Thus, we can use a JSCC codec that can adapt to a wide range of SNR conditions. During training, we sample the SNR uniformly between 0 dB and 20 dB. 
Due to the noise in the AWGN channel, the experiment results have small fluctuations. Therefore, we use the average value of five experiments as the final result.

\subsection{Performance Evaluation and Analysis}

\begin{figure*}[t]
	\centering
	\includegraphics[width=0.6\linewidth]{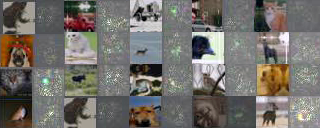}
	\caption{Gradient visualization of images.}    
	\vspace{-3ex}
	\label{figure_gradient}
\end{figure*}
\begin{figure}[t]
	\centering
	\includegraphics[width=0.7\linewidth]{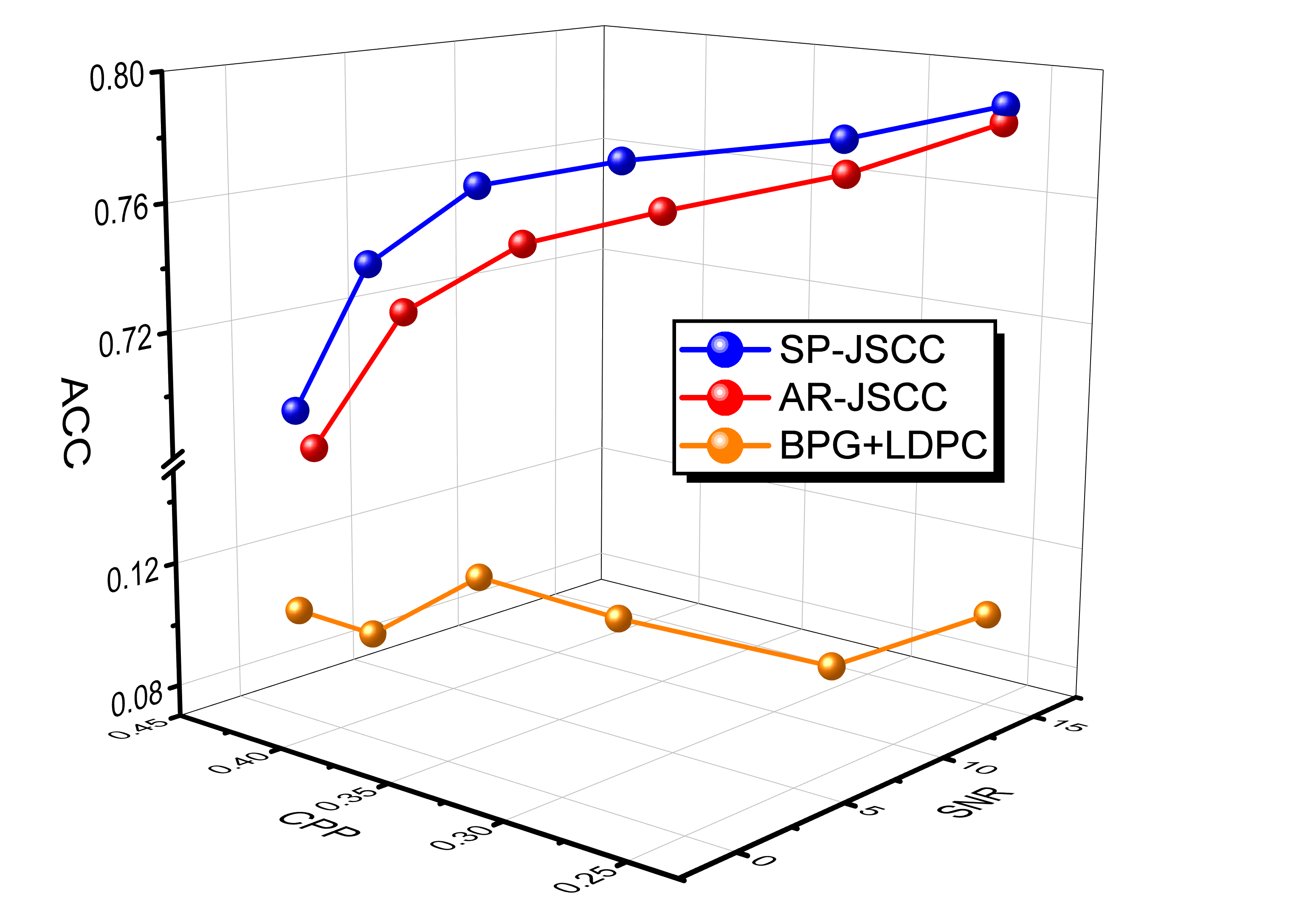}
	\caption{Graph of ACC versus SNR and CPP.}    
	\label{acc}
\end{figure}
\begin{figure}[t]
	\centering
	\includegraphics[width=0.7\linewidth]{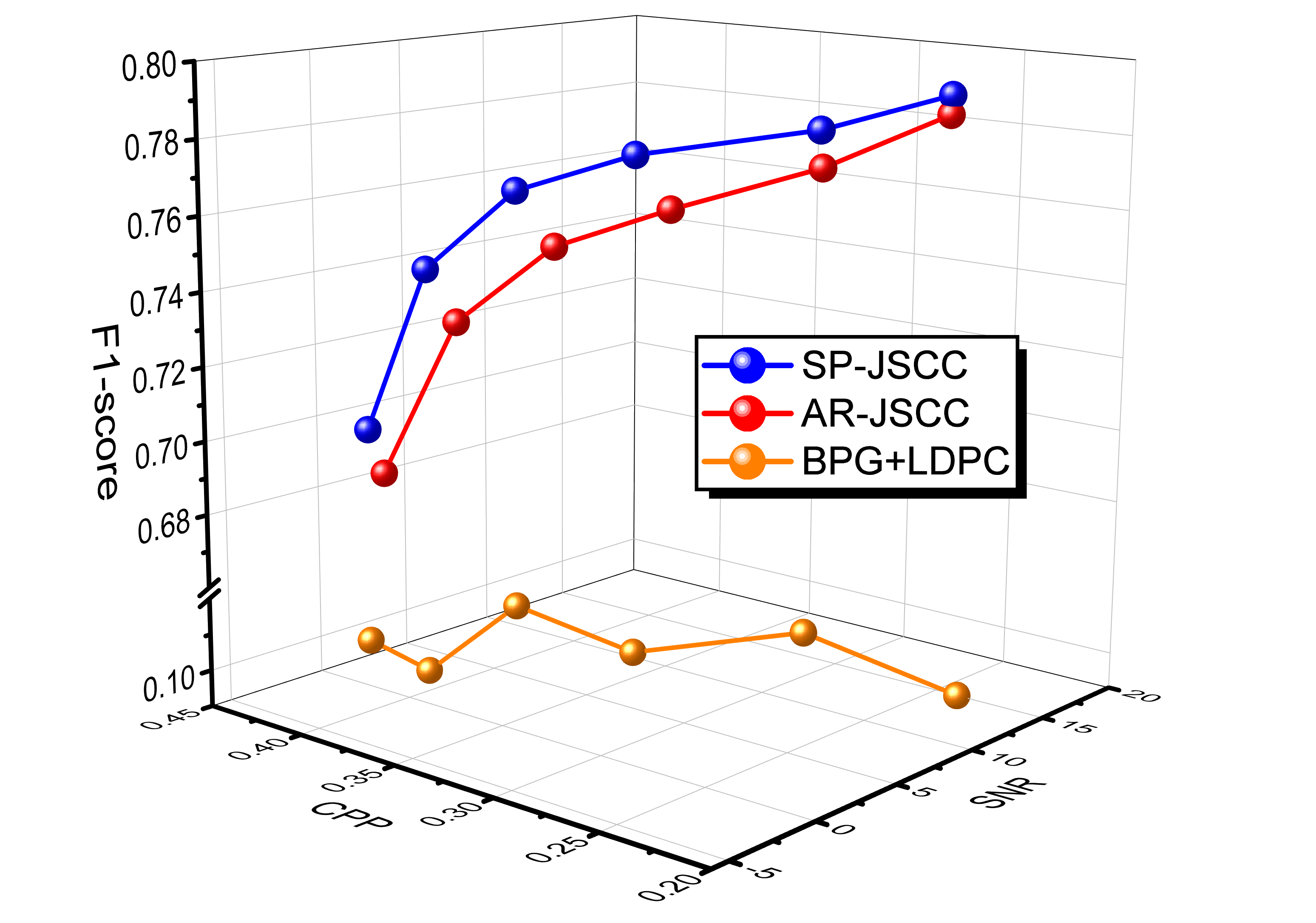}
	\caption{Graph of F1-score versus SNR and CPP.}    
	\vspace{-3ex}
	\label{f1score}
\end{figure}

\begin{figure}[t]
	\centering
	\includegraphics[width=0.7\linewidth]{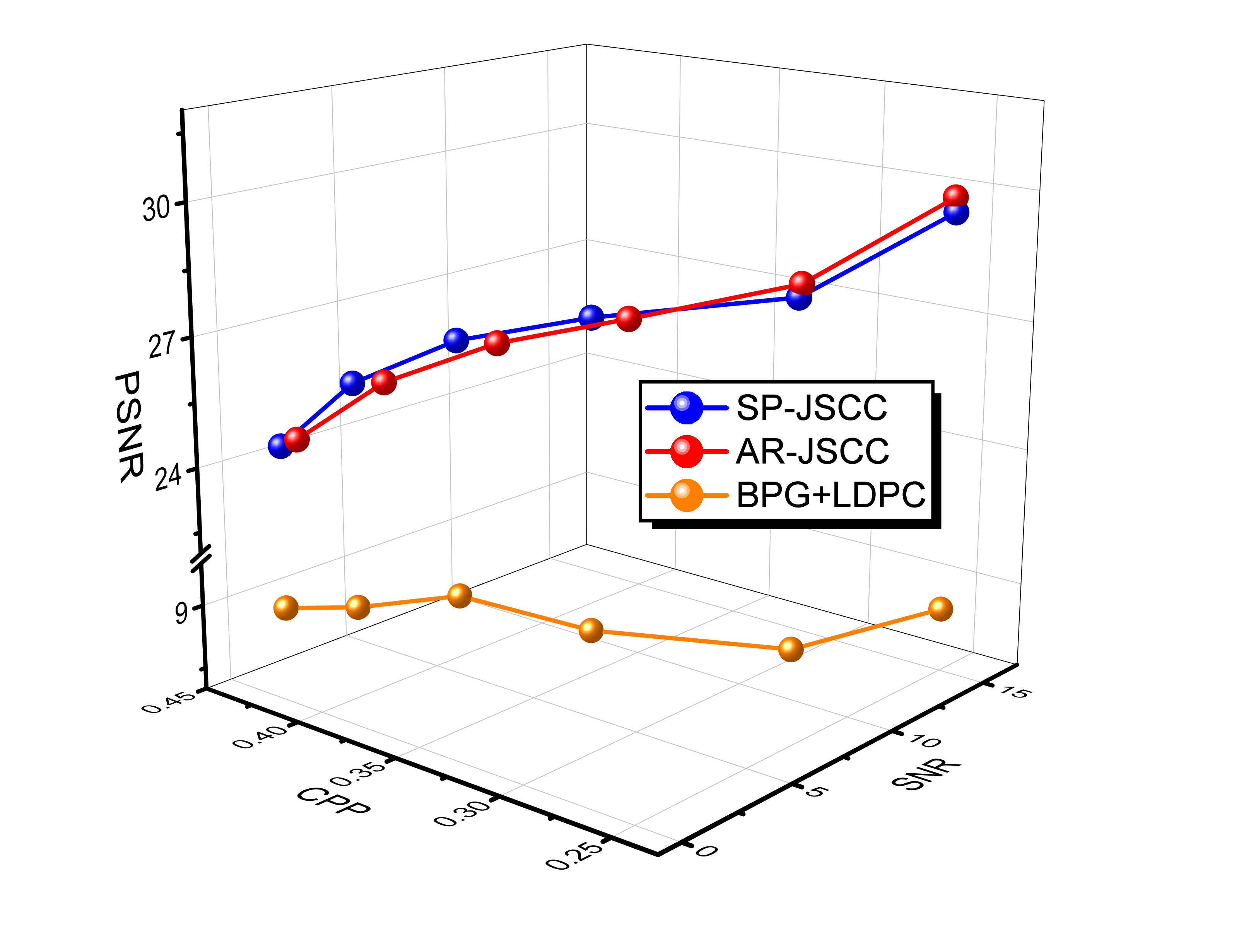}
	\caption{Graph of PSNR versus SNR and CPP.}    
	\vspace{-3ex}
	\label{psnr}
\end{figure}
\begin{figure}[t]
	\centering
	\includegraphics[width=0.7\linewidth]{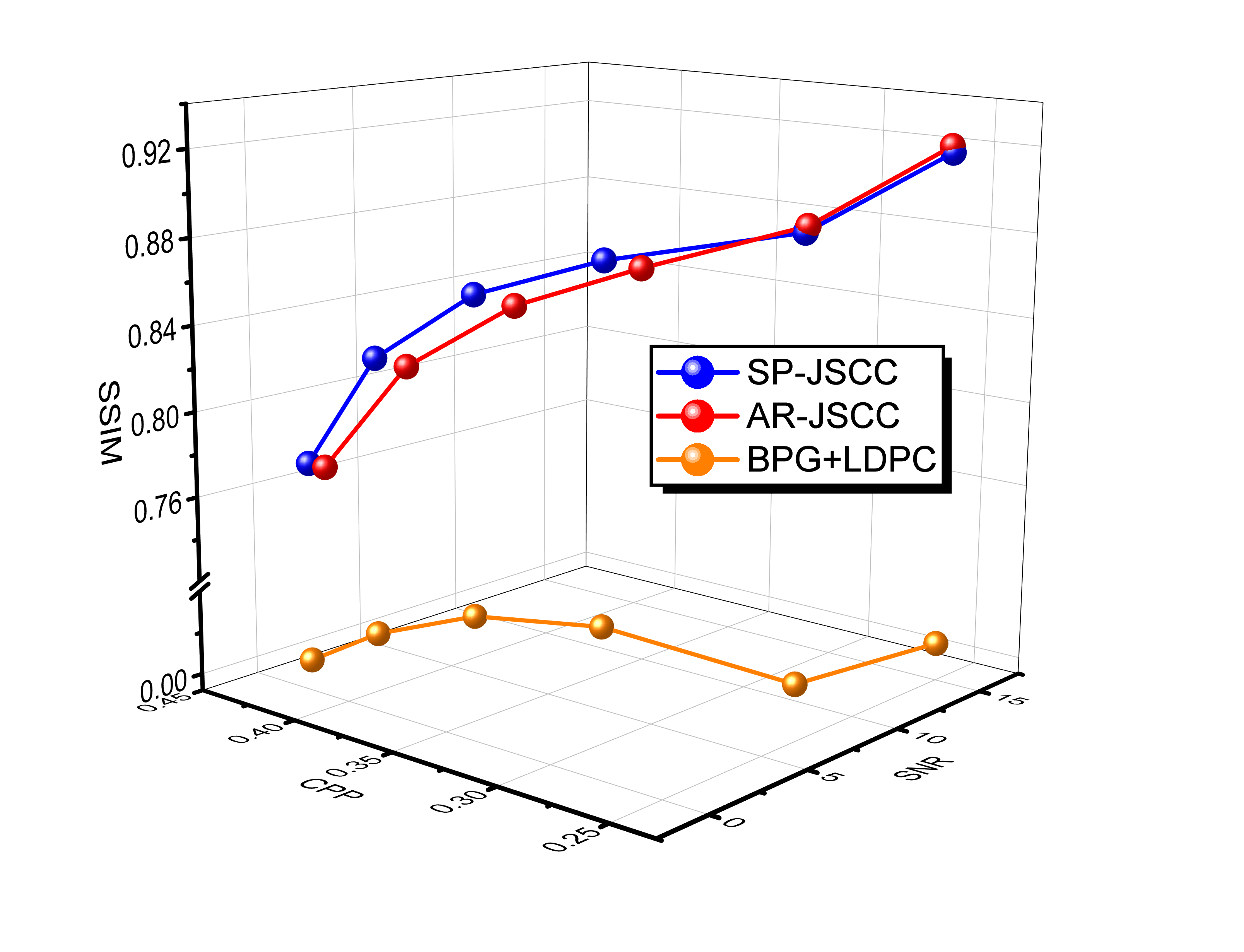}
	\caption{Graph of SSIM versus SNR and CPP.}    
	\vspace{-3ex}
	\label{ssim}
\end{figure}

\begin{figure}[htbp]
    \centering
    \subfigure[]{
        \includegraphics[width=2.5in]{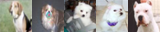}
        \label{origin}
    }
    \quad    
    \subfigure[]{
    	\includegraphics[width=2.5in]{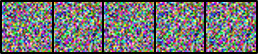}
        \label{label_for_cross_ref_3}
    }
    \quad    
    \subfigure[]{
    	\includegraphics[width=2.5in]{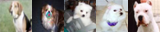}
        \label{label_for_cross_ref_3}
    }
    \quad    
    \subfigure[]{
    	\includegraphics[width=2.5in]{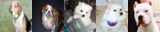}
        \label{label_for_cross_ref_3}
    }
    \caption{Reconstruction images at the receiver under 5 dB. (a) Original images, and the ground-truth is dog. (b) Reconstruction images of BPG, and perception results from left to right are bird, car, cat, deer, cat. (c) Reconstruction images of AR-JSCC, and perception results from left to right are cat, cat, deer, cat, horse. (d) Reconstruction images of SP-JSCC, and perception results are dogs. }
    \vspace{-2.5ex}
    \label{images}
\end{figure}

Fig \ref{figure_gradient} compares the original images and the corresponding SP-based weights. 
The object part in the image is more semantically important than the background part. As shown in Fig \ref{figure_gradient}, SP-based weights give the object part greater weight than the background part, that is, the distribution of SP-based weights pays more attention to the semantic part in the image. This shows the rationality of semantic weights.
Therefore, the SP-JSCC approach using the SP-based weights can enable the network focus on semantic consistency during transmission, rather than pixel-level consistency. 

Performances are related to transmission rate and channel conditions. The larger the transmission rate, the better the channel conditions, the more favorable the signal transmission, and thus the better performances. We next explore how different performances vary with transmission rate and channel conditions.
Fig \ref{acc} and Fig \ref{f1score} gives the variation curve of ACC and F1-score with transmission rate (CPP) and channel conditions (SNR) of different methods, respectively.
As shown in Fig \ref{acc} and Fig \ref{f1score}, ACC and F1-score are more competitive using SP-JSCC.
For example, ACC is improved by 1.38\% and 66\% compared with the AR-JSCC method and SSCC method (BPG+LDPC) at 5 dB, respectively. 
This is due to the SP-based loss function designed by SP-JSCC, which can retain semantic information that is beneficial to downstream AI tasks, and deserves a competitive ACC. It is worth mentioning that the BPG method cannot recover image under the same transmission rate and channel condition, and its performance is kept at the lowest value.
At the same time, ACC and F1-score are close to each other, which is due to the balanced distribution of categories on dataset.

Fig \ref{psnr} and Fig \ref{ssim} compares PSNR and SSIM of different methods, respectively, which can illustrate pixel-level consistency. As shown in Fig \ref{psnr} and Fig \ref{ssim}, PSNR and SSIM are comparable between SP-JSCC and AR-JSCC. This shows SP-JSCC does not compromise the pixel's information of images. In addition, BPG is poor recovery and low performance under the same transmission rate and channel condition.

Fig \ref{images} compares the image at the receiver of different methods under 5 dB. As shown in Fig \ref{images}, the visual quality of SP-JSCC is as good as AR-JSCC. There are subtle differences on images obtained by AR-JSCC and SP-JSCC, but images obtained by SP-JSCC can get correct perception results while AR-JSCC cannot. This shows that pixel-level consistency cannot guarantee the consistency of downstream AI task's perception results, and some tiny distortions at the pixel level may cause false perception results. In addition, the distortion of BPG is serious, and images cannot be restored.

The above experimental results show that, on one hand, SP-JSCC considers semantic information of the downstream AI task, which can improve the task performance. On the other hand, SP-JSCC considers the pixel's information of images and has high visual quality. When the image at the receiver not only needs to be understood by humans, but also needs to complete downstream AI tasks, SP-JSCC shows obvious superiority.

\section{conclusion}

To preserve semantic information at the receiver during the wireless image transmission, we propose SP-JSCC method, and give its network design and algorithm. 
SP-JSCC method quantizes semantic importance of pixels using gradients, and uses semantic distortion to guide the training process. 
The experimental results show that the SP-based weights focus on the object part of the image, and can effectively represent the semantic importance of pixels.
The improvement on the performance of downstream AI tasks is obtained by SP-JSCC, which can illustrate
the semantic information needed by the downstream AI task is retained.
SP-JSCC can be used for a wide range of AI tasks with no additional inference overhead. However, the disadvantage is that SP-JSCC increases the training overhead. In the future, we will consider other downstream AI tasks, and generalize it to high-resolution images. 



\bibliographystyle{unsrt}
\bibliography{conference_101719}

\end{document}